\newcommand{\smo}{Smol\v{c}i\'{c} }
\documentclass{PoS}

\usepackage{graphicx}
\usepackage{multirow}
\usepackage[comma]{natbib}
\title{Star-forming galaxies versus low- and high-excitation radio AGN in the VLA-COSMOS 3GHz Large Project}
\ShortTitle{SF galaxies vs. radio AGN in VLA-COSMOS 3GHz Large Project}
\author{\speaker{N. Baran}$^1$, V. Smol\v{c}i\'{c}$^1$, M. Novak$^1$, J. Delhaize$^1$, I. Delvecchio$^1$, P. Capak$^2$, F. Civano$^3$, N. Herrera-Ruiz$^4$, O. Ilbert$^5$, C. Laigle$^6$, S. Marchesi$^7$, H. J. McCracken$^6$, E. Middelberg$^4$, M. Salvato$^8$, E. Schinnerer$^9$\\
        $^*$E-mail: \email{nbaran@phy.hr}}
\abstract{We study the composition of the faint radio population selected from the VLA-COSMOS 3GHz Large Project, a radio continuum survey performed at 10~cm wavelength. The survey covers the full 2 square degree COSMOS field with mean $rms\sim2.3~\mu$Jy/beam, cataloging 10,899 source components above $5\times~rms$. By combining these radio data with UltraVISTA, optical, near-infrared, and Spitzer/IRAC mid-infrared data, as well as X-ray data from the \textit{Chandra} Legacy, \textit{Chandra} COSMOS surveys, we gain insight into the emission mechanisms within our radio sources out to redshifts of $z\sim5$. From these emission characteristics we classify our souces as star forming galaxies or AGN. Using their multi-wavelength properties we further separate the AGN into sub-samples dominated by radiatively efficient and inefficient AGN, often referred to as high- and low-excitation emission line AGN. We compare our method with other results based on fitting of the sources' spectral energy distributions using both galaxy and AGN spectral models, and those based on the infrared-radio correlation. We study the fractional contributions of these sub-populations down to radio flux levels of $\sim$10~$\mu$Jy. We find that at 3~GHz flux densities above $\sim$400~$\mu$Jy quiescent, red galaxies, consistent with the low-excitation radio AGN class constitute the dominant fraction. Below densities of $\sim$200~$\mu$Jy star-forming galaxies begin to constitute the largest fraction, followed by the low-excitation, and X-ray- and IR-identified high-excitation radio AGN.}
\author{\\\tiny$^1$ University of Zagreb, Physics Department, Bijeni\v{c}ka cesta 32, 10000 Zagreb, Croatia\\
	$^2$ Spitzer Science Center, California Institute of Technology, 220-6, Pasadena, CA, USA, 91125\\
	$^3$ Yale University, Department of Physics, 217 Prospect Street, New Haven, CT 06511-8499USA\\
	$^4$ Ruhr-Universit\"{a}t Bochum, Astronomisches Institut, NA 7/73, Universit\"{a}tsstr. 150, 44801 Bochum, Germany\\
	$^5$ Laboratoire d'Astrophysique de Marseille, P\^{o}le de l'\`{E}toile Site de Ch\^{a}teau-Gombert 38, rue Fr\'{e}d\'{e}ric Joliot-Curie 13388 Marseille, France\\
	$^6$ Institut d'Astrophysique de Paris, 98bis Boulevard Arago, 75014 Paris, France\\
	$^7$ INAF - OABO, Via Ranzani 1, 40127, Bologna, Italy\\
	$^8$ Max Planck Institute for Extraterrestrial Physics, Giessenbachstr. 1, 85748 Garching, Germany\\
	$^9$ Max Planck Institute for Astronomy, K\"onigstuhl 17, D-69117 Heidelberg, Germany
}

\FullConference{EXTRA-RADSUR2015 (*)\\
		20--23 October 2015\\
		Bologna, Italy

                \bigskip
                \hrule
                \bigskip

                \textnormal{(*) This conference has been organized
                  with the support of the Ministry of Foreign Affairs
                  and International Cooperation, Directorate General
                  for the Country Promotion (Bilateral Grant Agreement
                  ZA14GR02 - Mapping the Universe on the Pathway to
                  SKA)}
}
\begin{document}
\section{Introduction}
\label{sec:intro}
In extragalactic radio (cm) continuum surveys two main populations of sources are usually detected: star-forming galaxies and active galactic nuclei (AGN) (e.g., \citealt{miley80}, \citealt{condon92}, and recently \citealt{smolcic08}). A large fraction of radio emission originates from synchrotron radiation emitted by supernovae explosions, and from the vicinity of supermassive black holes \citep{condon92}. A smaller fraction arises from the free~-~free scattering processes. AGN emit throughout the electromagnetic spectrum, and not only are they detectable at the radio wavelengths, but certain types of AGN are powerful radio sources. In recent years, the presently favored classification scheme for radio sources focuses on their physical properties, and is linked to the existence (or the lack) of high-excitation optical emission lines (\citealt{hine79}, \citealt{hardcastle06, hardcastle07}, \citealt{allen06}, \citealt{smolcic08, smolcic09}, \citealt{heckman14}). Optically defined blue and green galaxies are the usual hosts of high-excitation radio AGN (HERAGN) which can also be identified via their optical, X-ray and/or mid-infrared photometry (e.g., \citealt{hardcastle13}). Low-excitation radio AGN (LERAGN) are only detectable at radio wavelengths, and are thought to be hosted by quiescent red galaxies (see \citealt{smolcic09}), and are also regarded as radio loud (\citealt{condon92}, \citealt{padovani15}). These two distinct types of radio AGN also exhibit differences in the way they accrete matter onto the central supermassive black hole (SMBH) (e.g., \citealt{hardcastle06}, \citealt{smolcic09}). HERAGN are found to be accreting radiatively efficiently, likely through the presence of an accretion disk, as postulated by the unified model of AGN (e.g., \citealt{urry95}). LERAGN are thought to be powered via advection dominated accretion flow (see \citealt{heckman14}), and do not fit into the unified model of AGN (e.g., \citealt{smolcic09}, \citealt{best12}). 
A set of proxies reflecting the physical properties of the sources detected in deep sky surveys, such as COSMOS, would allow for their separation into the above-mentioned sub-populations. In this work, we use proxies derived from optical, infrared, and X-ray data, to determine the nature of the sources detected in the VLA-COSMOS 3GHz Large Project (\smo et al., in prep), and to study the population mix of microjansky radio sources up to $z\sim5$. We separate the radio population into three main types: star-forming galaxies, and two types of AGN based on their X-ray, infrared emission, and host galaxy colors. In Section~\ref{sec:data} we present the dataset. The classification of the radio source components is described in Section~\ref{sec:counterparts}. The results of the separation, and their interpretation in the context of the sources' physical properties follow in Section~\ref{sec:method}. Finally, in Section~\ref{sec:millijansky} we discuss the composition of the faint radio source population.
\section{Data}
\label{sec:data}
\subsection{Radio Data}
The VLA-COSMOS 3GHz Large Project encompasses 384 hours of observations with the Karl J. Jansky Very Large Array (VLA) in S-band (centered at 3~GHz with 2,048~MHz bandwidth) as described in detail in \smo et al. (in prep). The source component catalog was extracted from the 2.6~deg$^2$ 3~GHz map, and contains 10,899 components with a local signal-to-noise ratio (SNR)$\geq5$. Within the 2~deg$^2$ of the COSMOS field, the median $rms$ is $\sim2.3~\mu$Jy/beam, at a resolution of $0.75''$. 
\subsection{Mid-infrared - near-ultraviolet data and redshifts}
\label{subsec:mirdata}
\begin{figure}
	\begin{minipage}[l]{7.6cm}
			\includegraphics[bb = -63 70 865 856, width=1.0\linewidth]{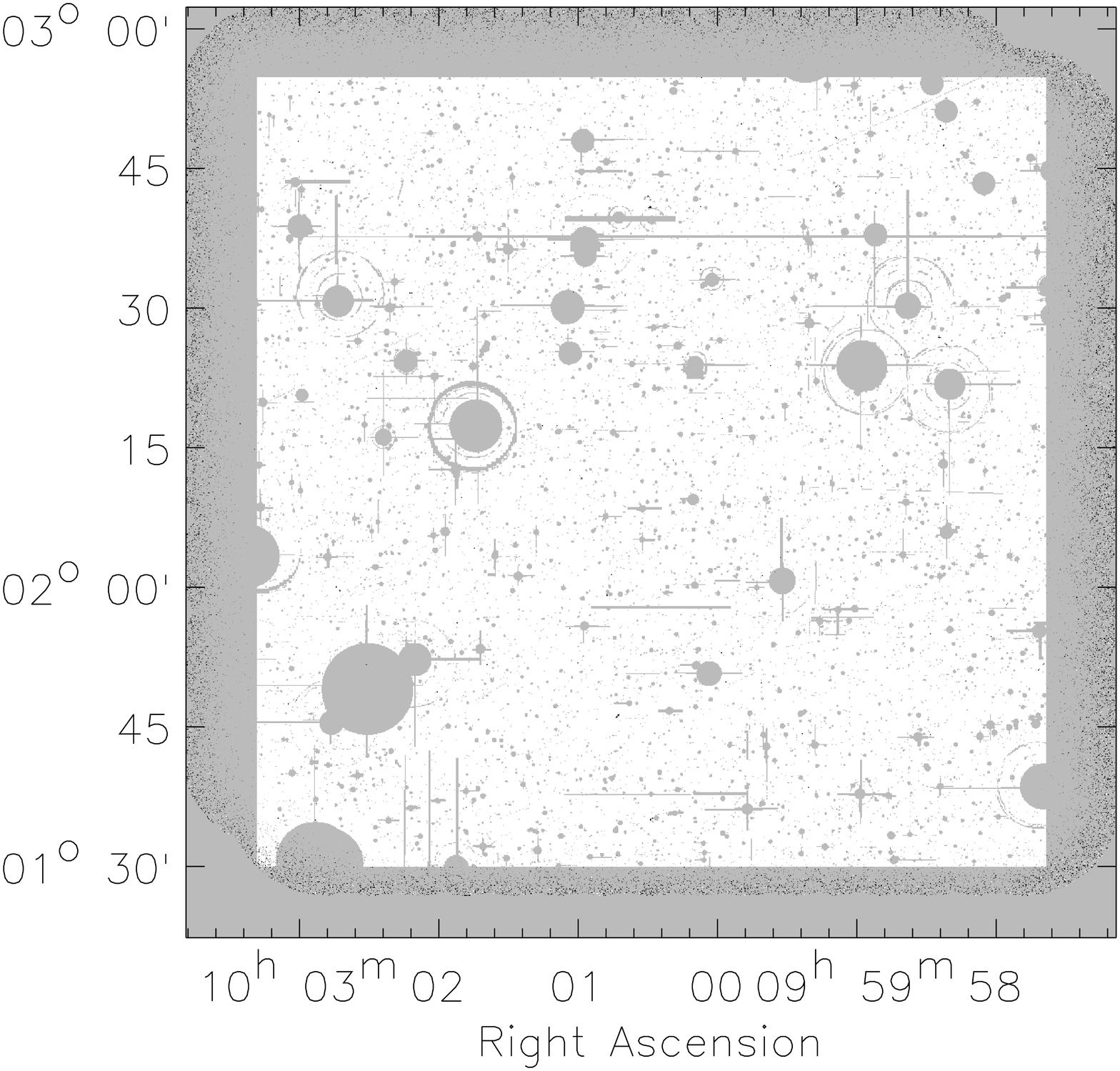}
			\caption{VLA-COSMOS 3GHz Large Project mosaic with regions masked due to saturation from stars and map edges in COSMOS2015 catalog grayed-out.}
			\label{fig:coverage}
	\end{minipage} \ \hfill \ 
	\begin{minipage}[l]{6.9cm}
			\includegraphics[bb=95 70 765 850,width=0.8\linewidth]{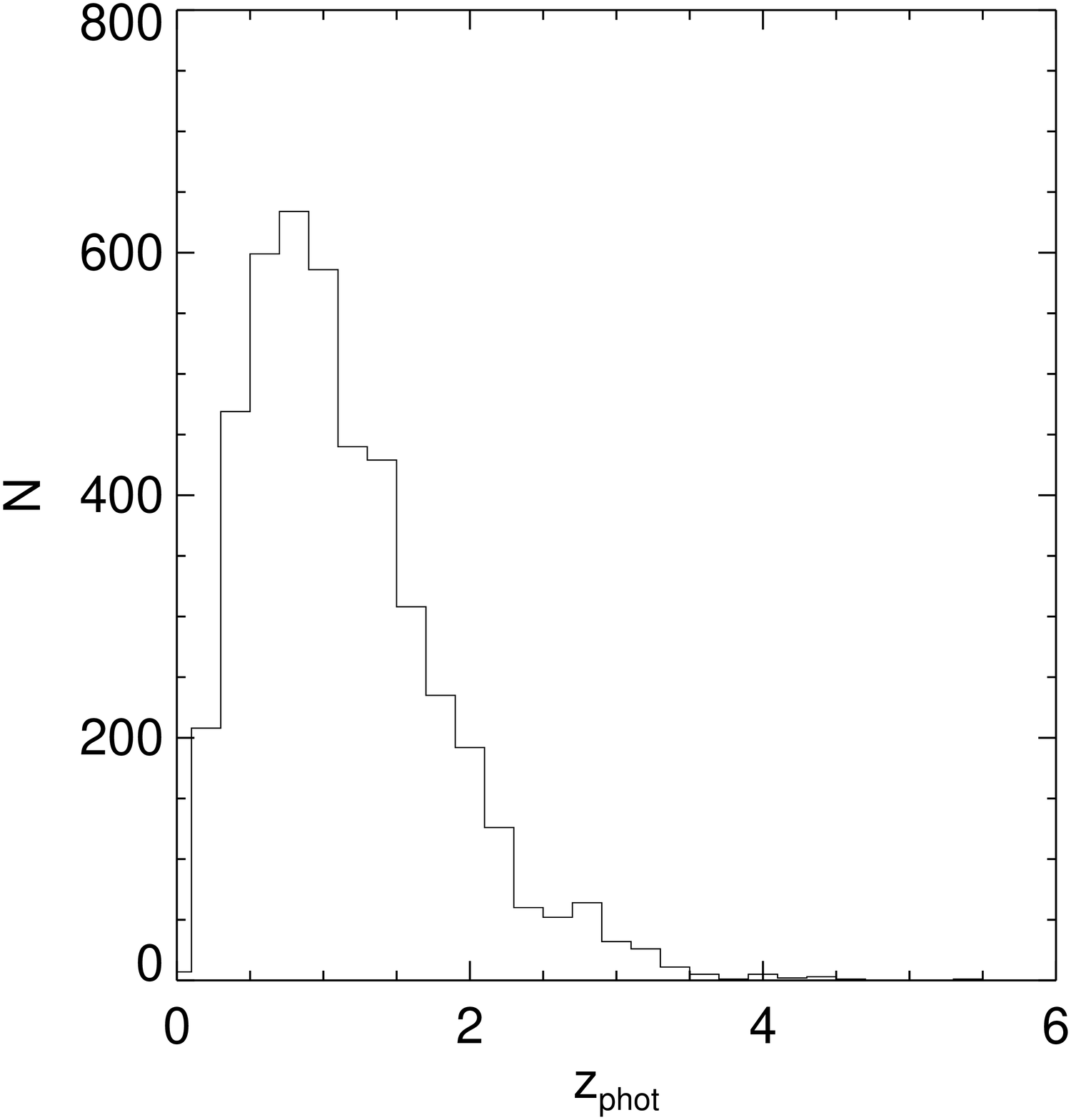} 
			\caption{Photometric redshift distribution of the radio-COSMOS2015 matched sample of 6,214 sources.}
			\label{fig:ZphotZspec}
	\end{minipage}
\end{figure}

We here use the photometry and photometric redshift catalogs (COSMOS2015 hereafter) described by Laigle et al. (submitted). These contain 1,182,108 sources detected in the map resulting from the chi-squared sum of $YJHK_S$ and Subaru $z^{++}$ images, where $YJHK_S$ is taken from UltraVISTA DR2\footnote{A detailed description of the survey and data products can be retrieved at: http://ultravista.org/release2/uvista\_dr2.pdf}. The COSMOS2015 catalog lists optical and NIR photometry in over 30 bands for each source. Laigle et al. have computed photometric redshifts for the sources using \textsc{LePhare} (see also \citealt{ilbert09, ilbert13}). Using the full COSMOS spectroscopic data, they report a photometric redshift accuracy of $\sigma_{\Delta z/(1+z_s)}~=~0.007$ with respect to zCOSMOS \citep{lilly07}, and $\sigma_{\Delta z/(1+z_s)}~=~0.021$ for $3\textless~z\textless6$ (cf. Table~4 in Laigle et al., submitted). The VLA-COSMOS 3GHz Large Project mosaic with overlaid masked areas (due to saturation from stars, and map edges) in the optical-NIR data is shown in  Fig.~\ref{fig:coverage}. The resulting effective area with accurate UV-NIR photometry is 1.8~deg$^2$.

We use the latest spectroscopic redshift catalog available to the COSMOS team (M.~Salvato in prep.). The catalog compiles a range of spectroscopic surveys of the COSMOS field, including SDSS (DR12; \citealt{alam15}), zCOSMOS (PI: S. Lilly), VIMOS Ultra Deep Survey (VUDS; \citealt{lefevre15}), MOSFIRE (Scoville et al. 2015 in prep.), MOSDEF \citep{kriek15}, and DEIMOS\footnote{DEIMOS PI: Jehan Kartheltepe, Peter Capak, Mara Salvato, G\"{u}nther Hasinger and Nick Scoville}.
\subsection{X-ray data}
\label{subsec:xraydata}
We use the most recent COSMOS X-ray catalog of point sources \citep{marchesi16}, drawn from the \textit{Chandra} COSMOS-Legacy survey \citep{civano16}. The \textit{Chandra} COSMOS-Legacy survey combines the \textit{Chandra} COSMOS \citep{elvis09} data with the new \textit{Chandra} ACIS-I data (2.8~Ms observing time) resulting in a total exposure time of 4.6~Ms over 2.15~deg$^2$ area, reaching a [0.5~-~2]~keV limiting depth of 2.2$\times$10$^{-16}$~erg~s$^{-1}$~cm$^{-2}$, and 8.9$\times$10$^{-16}$~erg~s$^{-1}$~cm$^{-2}$ in the [0.5~-~10]~keV band \citep{civano16}.
The catalog contains 4,016 X-ray point sources in the \textit{Chandra} COSMOS-Legacy survey, out of which 3,877 have optical--MIR counterpart \citep{marchesi16}. For these latter, spectroscopic (see \citealt{civano16}), or photometric redshifs (Salvato et al. in prep.) are available. The catalog lists X-ray fluxes, and intrinsic (i.e. unobscured) X-ray luminosities in the full [0.5~-~10]~keV, soft [0.5~-~2]~keV, and hard [2~-~10]~keV X-ray bands.

\section{Multi-wavelength counterparts of VLA-COSMOS 3GHz Large Project source components}
\label{sec:counterparts}
\subsection{MIR - Optical counterparts}
Out of 10,899 radio source components in the 3~GHz map, 8,750 (80\%) fall within the unmasked 1.8~deg$^2$ area in the optical--NIR maps (see Fig.~\ref{fig:coverage}, and Sec.~\ref{sec:data}), where the most accurate photometry and photometric redshifts are available. We then positionally match those radio source components with the sources listed in the COSMOS2015 catalog using a 0.4$''$ matching radius. This yields 7,122/8,750 (81\%) matches. We estimate a false match probability of $\lesssim~2$\% based on a positional match of 10,899 randomly positioned sources. We have verified that photometric redshift accuracies for our radio-COSMOS2015 matched sources ($\sigma_{\Delta z/(1+z_s)}=0.01$) are in agreement with the accuracies given by Laigle et al. (see Sec.~\ref{subsec:mirdata}). To select only sources with the most accurate photometric redshifts, we have introduced a cut in $i$-band magnitude of $m_i\le25.5$ \citep{ilbert09}. This reduces our sample to 6,214 radio sources with MIR-optical counterparts. Spectroscopic redshifts are available for 2,553 (41\%) sources, while 3,661 (59\%) have only photometric redshifts. The photometric redshift distribution of the matched sources is shown in the Fig.~\ref{fig:ZphotZspec}.
\subsection{X-ray counterparts}
\label{sec:xraycounterparts}
Of the 6,214 sources from the radio~-~COSMOS2015 dataset, 816 (13\%) are associated with the X-ray point sources (see Sec.~\ref{subsec:xraydata}) according to the X-ray~-~optical~-~MIR matched catalog of
 \citet{marchesi16}.
\section{Radio source classification}
\label{sec:method}
\subsection{X-ray AGN}
\label{subsec:xrayagn}
We consider all sources in our sample with intrinsic X-ray full band [0.5~-~8]~keV luminosity ($L_X$) above $log(L_X/\mathrm{erg~s}^{-1})=42$ to be X-ray AGN (e.g., \citealt{szokoly04}). From a total of 6,214 sources, 767 ($\sim13\%$) are selected as X-ray AGN. This amounts to 94\% of the full radio-X-ray sample, as defined in Section~\ref{sec:xraycounterparts}.
\subsection{Mid-infrared AGN}
\label{subsec:miragn}
To complement the X-ray AGN selection criterion, we apply the MIR-based selection method of \cite{donley12}. This method identifies the warm dusty tori surrounding the central SMBH, using the range of colors constructed from four IRAC bands at 3.6, 4.5, 5.8 and 8~$\mu$m (see \citealt{donley12} for details). Using this method, we find 365 MIR-identified AGN in total, out of which 201 were also identified as X-ray AGN. Although $\sim$18\% of the sources classified as X-ray AGN also satisfy MIR AGN criteria, $\sim3\%$ of sources in our sample fit MIR AGN criterion only. Thus, both MIR and X-ray AGN selection criteria are required to reach a more complete perspective of radiatively efficient AGN activity in our radio sample.
\subsection{Rest-frame optical colors as a proxy for star-forming and quiescent galaxies}
We find that in 4,431 ($\sim85\%$) of the 6,214 radio-selected sources in our sample do not display signatures of AGN activity in the X-ray or MIR. We therefore assume that emission of the host galaxy (and not that of AGN) dominates their optical-MIR spectrum of these objects. To separate galaxies based on their extent of star-formation activity, we use a rest-frame optical color-based separation method, already used and tested by \cite{ilbert10}. This method separates sources based on the their near ultraviolet ($NUV$) and $r$ rest-frame colors ($M_{NUV} - M_{r}$) corrected for internal dust extinction (see \citealt{ilbert10} for details). Sources are considered \textit{quiescent} (red) if $M_{NUV}~-~M_{r}>3.5$, having \textit{intermediate} star-formation activity (green) if $3.5>M_{NUV}~-~M_{r}>1.2$, and having \textit{high} star-formation activity (blue) if $M_{NUV}~-~M_{r}<1.2$. By applying the method to the 4,431 sources that were not selected as X-ray or MIR AGN, we find 1,595 \textit{intermediate}, and 2,836 \textit{high-activity} galaxies, which, as shown below can be considered as star-forming galaxies. Further, we find 852 red, quiescent galaxies.
\subsection{Physical properties of differently selected radio source populations}
Radiation arising from AGN carries a signature of physical mechanisms at play. In the absence of spectral information which would directly prove the (non)existence of high-excitation optical emission lines, we assess the radiative efficiency of accretion using proxies derived from X-ray, MIR and rest-frame optical emission. We thereby separate AGN population into HERAGN dominated, and LERAGN dominated sub-populations. 

HERAGN are thought to have radiatively efficient modes of accretion (see \citealt{evans06}, \citealt{smolcic09},  \citealt{best12}). Signatures of radiatively efficient accretion include a powerful X-ray emission, and MIR emission if a warm dusty torus is present around the central SMBH \citep{donley12}. Thus, we expect the X-ray, and MIR identified AGN sources to be predominantly consistent with properties of HERAGN. 

LERAGN are radiatively inefficient AGN \citep{hine79}, and do not display properties consistent with the unified model of AGN \citep{smolcic09}, remaining mostly undetected in X-ray and MIR. In the local Universe, LERAGN are hosted by optically red galaxies (\citealt{evans06}, \citealt{smolcic09}, see also \smo 2016, this Volume). Since the red rest-frame color indicates a low level of star-formation, consistent with quiescent galaxies, the radio emission from non-HERAGN radio sources with red rest-frame optical colors likely arises from AGN activity. Thus, we expect that red non-HERAGN radio sources are predominantly LERAGN.

For radio sources that do not display characteristics of AGN, we assume most of radio emission is arising from the supernovae explosions within the galaxies, linked to their star-formation rate \citep{ulvestad82}. Therefore we classify as a non-active star-forming galaxy any radio source in our sample that displays green or blue optical colors (hence has intermediate to high levels of star formation activity) and is not previously identified as LERAGN or HERAGN. This has been verified by Delhaize at al. (in prep.) who find that such objects closely follow the infrared-radio correlation of star-forming galaxies.

We have compared our classification with the results based on the spectral energy distribution (SED) fitting, performed on the same sample by I.~Delvecchio et al. (in prep.). They use an updated set of \textsc{Magphys} templates suitable to fit the SED of galaxies hosting an AGN as well as regular galaxies (\citealt{dacunha08}, \citealt{berta13}). In addition to the X-ray and MIR-selected AGN, as defined here, they identify further AGN, based only on the SED fitting results \citep{delvecchio14}. A comparison with their method suggests that our X-ray+MIR-AGN sample is about 70\% complete. This is consistent with the findings based on Very Large Baseline Array (VLBA) (PI:~E.~Middleberg, see Herrera-Ruiz et al. in prep.) point source catalog at 1.4~GHz. Out of 468 VLBA sources in total, we find 301 positional matches with our radio-COSMOS2015 sample in a 0.1$''$ radius, where 203 (67\%) have already been selected as AGN. Since VLBA point sources are very compact ($\sim$1/100$''$ resolution) they can be considered either as compact starburst regions or AGN. The SFR analysis points to the latter as the source of the radio emission (e.g., Delvecchio et al., in prep). The lack of AGN features from SED fitting further suggests the predominantly lower radiative efficiency of our red galaxy (LERAGN dominated) sample, with respect to our X-ray and MIR-selected AGN. This is consistent with the results from Delhaize et al. (in prep.) who find a significant radio excess with respect to star-formation rate for our red, quiescent galaxy population, thus confirming its properties are dominated by LERAGN. A slight radio excess in HERAGN is also notable, while blue and green galaxies show no radio excess, and are, thus, considered predominantly star-forming. Our classification scheme is summarized in Table~\ref{tab:sampleCensus}. 
\begin{table}[]
	\centering
	\caption{Overview of the sample, the fraction of each type in our sample (6,214), and classification.}
	\begin{tabular}{c c c c c c}
		Sample group         & Fraction & Likely dominated by                                             \\
		\hline\hline         
		X-ray AGN          & 12\%                              & \multirow{2}{*}{HERAGN}                             \\
		MIR AGN            & 3\%                               &                                                   \\
		\hline               
		Red galaxies       & 14\%                              & LERAGN                                              \\
		\hline               
		Green galaxies     & 26\%                              & \multicolumn{1}{c}{\multirow{2}{*}{Star-forming}} \\
		Blue galaxies      & 45\%                              & \multicolumn{1}{c}{}                              \\
		\hline\hline             
	\end{tabular}
	\label{tab:sampleCensus}
\end{table}
\section{Composition of the sub-millijansky radio source population}
\label{sec:millijansky}
\begin{figure}
	\centering
	\includegraphics[bb = 0 60 1814 840, width=1\linewidth]{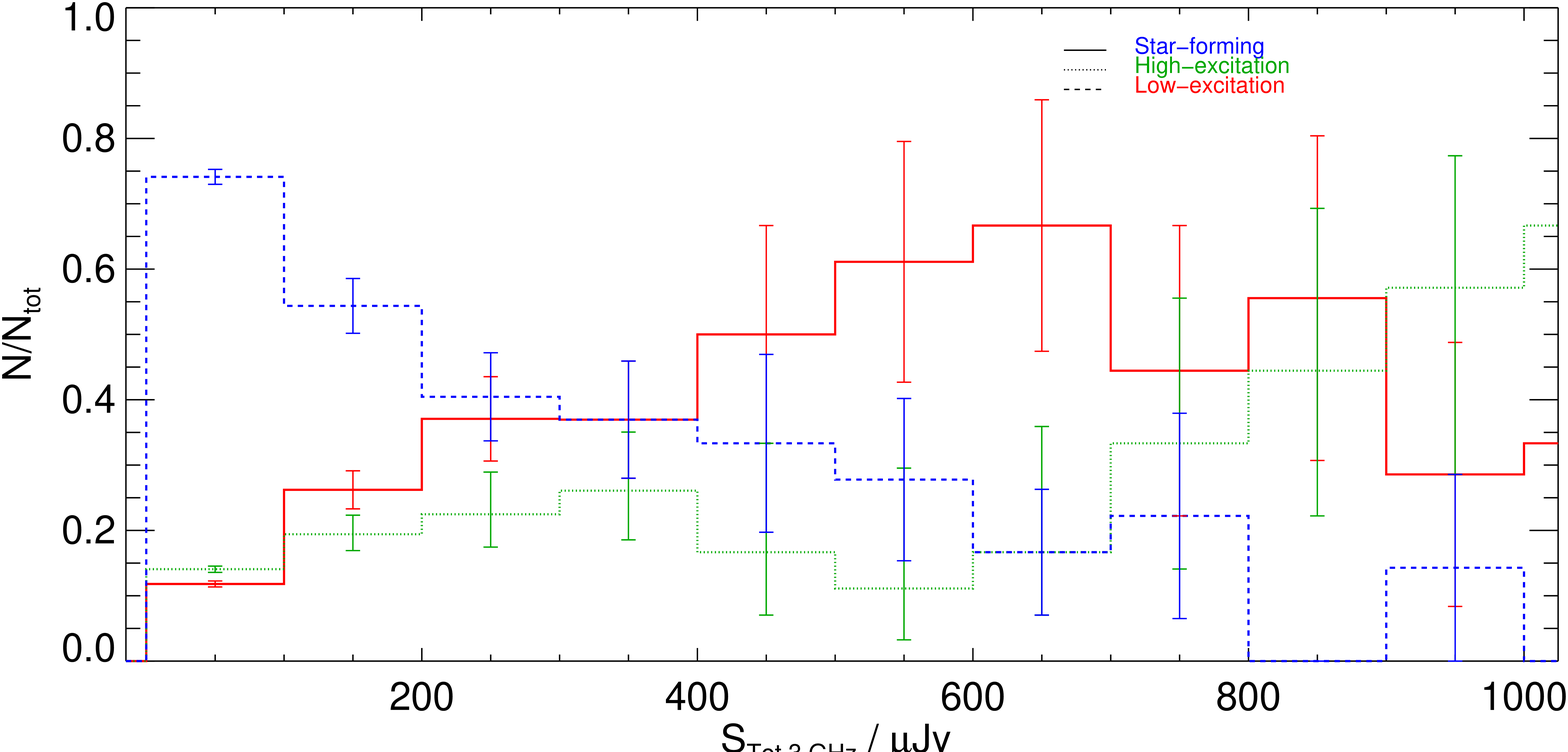}
	\caption{Fractional contribution of each sub-population to the total number of sources, as a function of 3~GHz radio flux.}
	\label{fig:frac}
\end{figure}

After separating the 3~GHz radio sources into three distinct categories, we study their fractional contributions to the total number of sources as a function of 3~GHz radio flux. This is shown in Fig.~\ref{fig:frac}. It is notable that above $\sim400\mu$Jy the LERAGN dominate, followed by HERAGN and star-forming galaxies. Below $\sim200\mu$Jy the situation reverses, and star-forming galaxies dominate the fraction while low-excitation sources contribute the least. This is consistent with previous results from \citealt{smolcic08}, \citealt{mignano08}, \citealt{bonzini13} and \citealt{padovani15}.

In summary, we have classified objects detected in the VLA-COSMOS 3GHz Large Project into sub-populations dominated by high- and low-excitation radio AGN, and star-forming galaxies, using their X-ray, MIR, and rest-frame optical colors. We have thereby examined the composition of the sub-millijansky radio source population as a function of 3~GHz flux.
\section{Acknowledgements}
\tiny{The authors of this work acknowledge the support of European Union's Seventh Framework program under grant agreement 333654 (CIG, 'AGN feedback'; N.B., V.S.), grant agreement 337595 (ERC Starting Grant, 'CoSMass'; V.S., J.D., M. N.), and Croatian science foundation HRZZ (N.B.).}

\end{document}